# Coefficient of Restitution based Cross Layer Interference Aware Routing Protocol in Wireless Mesh Networks

Sarasvathi V[1], Snehanshu Saha[2], N.Ch.S.N. Iyengar[3] and Mahalaxmi Koti[4]

[1,2,4]Department of Computer Science and Engineering, PESIT Bangalore South Campus, Bangalore-560 100, India.
[3]School of Computing Science & Engineering, VIT University, Vellore-632014, Tamilnadu, India.
sarasvathiram@gmail.com[1], snehanshu.saha@gmail.com[2], nchsniyengar48@gmail.com[3], nimishakoti@gmail.com

***Abstract***: In Multi-Radio Multi-Channel (MRMC) Wireless Mesh Networks (WMN), Partially Overlapped Channels (POC) has been used to increase the parallel transmission. But adjacent channel interference is very severe in MRMC environment; it decreases the network throughput very badly. In this paper, we propose a Coefficient of Restitution based Cross layer Interference aware Routing protocol (CoRCiaR) to improve TCP performance in Wireless Mesh Networks. This approach comprises of two-steps: Initially, the interference detection algorithm is developed at MAC layer by enhancing the RTS/CTS method. Based on the channel interference, congestion is identified by Round Trip Time (RTT) measurements, and subsequently the route discovery module selects the alternative path to send the data packet. The packets are transmitted to the congestion free path seamlessly by the source. The performance of the proposed CoRCiaR protocol is measured by Coefficient of Restitution (COR) parameter. The impact of the rerouting is experienced on the network throughput performance. The simulation results show that the proposed cross layer interference aware dynamic routing enhances the TCP performance on WMN.

***Keywords***: Coefficient of Restitution, Wireless Mesh Networks, Partially Overlapped Channels, Round Trip Time, Multi-Radio, Multi-Channel.

## 1. Introduction

Ever since the evolution of communication began, quality of service (QoS) has become imperative to be considered in computer networks. Nowadays, multimedia communication on the Internet has been dominant communication. When the number of users on the multimedia communication channel is increased or more traffic on the Internet, there may be packet loss and quality degradation. The emerging interactive applications like multimedia streaming and multiplayer games demand less round trip time, so RTT plays an important role in increasing the throughput. Our primary aim is to reduce the RTT, loss rate and collision, caused by interference, for refining TCP performance in WMN.

The IEEE 802.11 b/g network operates in the 2.4GHz ISM frequency and the frequency spectrum is divided into 11 channels, in which, three of the channels are non-overlapping or orthogonal channels such as channel 1, 6 and 11. Since, the number of orthogonal channels is limited; it is not possible to allocate channels for all the neighboring nodes. Moreover, if the same channel is assigned to more than one neighboring node, that will lead to co-channel interference in simultaneous transmission and finally, it results in throughput degradation. In the event of, one channel overlapping with another channel, for instance, Channel 1 overlaps with channel 3, they are called partially overlapping channels. Most of the existing system design considers POC as a danger because it severely affects the transmission between the nodes. An efficient channel assignment technique with POC [13] solves the interference problem and also produces significant improvement in parallel transmission and throughput.

Recently, Wireless Mesh Network has been an attractive technology platform for Internet service and it caters to diversified segments like academic, Industry and community networks [3]. The WMN provides seamless reliability, excellent coverage and high performance compared to the single hop networks. By exploiting the MRMC in WMN, greater network throughput can be achieved than the single radio single channel, due to the advantage of parallel transmissions. In multi-radio setup, each node is equipped with multiple radios and each radio is assigned to different channels for simultaneous transmission. Most of the research in WMN with IEEE802.11b/g standard, orthogonal channels is used. In this work, we primarily focus on POC, which increases the number of users accessing the Internet. But the major problem with POC is that the interference between the adjacent channels and its effect, as it reduces the network throughput badly.

### 1.1 Congestion in WMN

The densely deployed nodes in IEEE 802.11 WMN can cause network congestion that leads to a packet drop, delay in delivery and frequent disconnection. Generally, the data from the source is reached in the Internet through the gateway with multi hop access, so congestion occurs more near the gateway, but random at other network destinations. Most of the congestion control algorithms in the wired networks try to estimate the available capacity, i.e. bandwidth, queue size in the router, to fix the congestion window on the sender side. These congestion control algorithms do not apparently find the real congestion status of the wireless networks because of various reasons, such as channel interference, mobility and congestion. The various algorithms [16, 15, 1] have been developed for wireless networks for refining the ability of TCP to judge the congestion status more efficiently. These solutions are categorized into two kinds:

- **End-to-End congestion control method:** It reacts very slowly in wireless networks because of the waiting time for acknowledgement (ACK) is more.
- **Hop-by-Hop congestion control method:** It reacts quickly to detect the status of the link and intermediate nodes, and it can make decision effective.

The hop-by-hop delay is accumulated into the end-to-end delay, so controlling the single hop delay ensures that the less amount of end to end delay. Based on the channel access at each hop, the per- hop delay would significantly change.

In this work, the RTS/CTS scheme at the MAC layer is used to estimate the congestion status of the link and we propose a contention mechanism algorithm at the MAC layer, and then hop-by-hop RTT is estimated for dynamic routing. The performance of the algorithm is evaluated using the COR. The channel busy time and throughput is considered to measure the network, whether it is highly congested or not. Our simulation results show that the proposed method can yield less delay, good throughput and less packet loss to the interference situation. It can also provide QoS and minimize RTT along the path.

This paper is organized as follows: Section II describes the existing congestion control and routing algorithms in wireless networks. In Section III, the system model is explained, in which, contention algorithm in MAC layer is modified and the resultant RTT calculation is presented. The routing algorithm is explained in section IV. The simulation settings, graphs and performance evaluation using COR are analyzed in section V. The paper is concluded and the scope is discussed in section VI.

## 2. Related Work

Basic TCP congestion control does not perform well in the wireless networks because of the fact that difficult to differentiate between the congestion event and bit error event. In [6], an improved TCP congestion control Algorithm for wireless networks was proposed. The basic TCP congestion control algorithm is modified to enhance the performance of TCP in wireless networks. The multiplicative decrease is refined in TCP NewReno and the statistics counter is used to monitor the frequencies of timeout occurrences and 3 duplicate acknowledgements. The value of the counter and the quantum of time between two consecutive timeouts decide the congestion losses or bit error. This algorithm gives better performance in heterogeneous networks and modification has been done only on the sender side of TCP, no burden on the internal network. If there is a real congestion, then it performs as original TCP NewReno, otherwise it carries on transmitting at a good speed. So, the capacity of the network is utilized properly in the case of bit errors.

The packet arrival and departure time are compared, to distinguish between the congestion loss and error losses in Wireless TCP [10]. This is an end-to-end semantic mechanism used in Wireless TCP, and it does not half down its transmission rate like TCP, instead sending rate is adjusted at the receiver based on inter-packet delay metric. The WTCP uses the rate based transmission and the feedback is taken from the receiver to retransmit the packet.

The channel capacity is subject to fading, So, Hasen et al [18] presented closed form expression to improve the channel capacity, which increases the SNR level. The TCP Westwood [12] refines the TCP Reno for wireless networks and it primarily depends on end-to-end bandwidth estimation to find out the causes of packet losses. The inspection or interception of packets at proxy node is not required in TCP Westwood as it continuously monitors the ACK returning rate. The network capacity is calculated by measuring the arrival rate of ACK and the same is denoted by SBW[j] .Also, the smoothed value BWE[j] is calculated by low-pass filtering the sequence of SBW[j] .

$$ SBW[j] == \frac{Packet\ Size}{current\_time - prev\_ACK\_time} \quad (1) $$

$$ BWE[j] = \frac{(1-t)*(SBW[j]+SBW[j-1])}{2} + t*BWE[j-1] \quad (2) $$

Where $t$ is the low-pass filter factor, *packet_size* specifies packet's size, *current_time* is the most recent time, and *prev_ACK_time* is the time when ACK received. This method tries to estimate the approximate bandwidth to set the congestion window size.

The TCP Vega [8] uses the modified slow start mechanism and the new timeout mechanism for congestion avoidance. The objective of TCP Vega is to maintain the correct amount of data in the network. Based on the variation of estimated extra data present in the network, the algorithm decides the sending rate. If the source is transmitting too much of data, there will be a delay in getting the acknowledgement and it will lead to congestion. The TCP Vega finds the *BaseRTT* when the network is not congested, and in this case, the expected rate is given by

$$ Expected\ Rate = \frac{Congestion\ Window}{BaseRTT} \quad (3) $$

Where, the congestion window indicates the number of bytes in transition.

The current sending rate is calculated by actual RTT.

$$ Actual\ Rate = \frac{Congestion\ Window}{ActualRTT} \quad (4) $$

The difference between the *Actual Rate* and *Expected Rate* is calculated and accordingly the congestion window is adjusted.

$$ Diff = Expected\ Rate - Actual\ Rate \quad (5) $$

The thresholds $\alpha$ and $\beta$ are used to measure the amount of data present in the network. If Diff < $\alpha$, then the congestion window is increased linearly. When Diff > $\beta$, the congestion window is decreased linearly. If $\alpha$ < Diff < $\beta$, then the congestion window is unchanged. However, these congestion control algorithms [1,2,3,4,5] may not be appropriate for MRMC WMN with partially overlapping channels, where the packet loss is due to the interference and its dynamic nature of channel assignment. Therefore, instead of the typical congestion control algorithm, we propose a CoRCiaR protocol, which involves MAC and routing layers for reliable TCP protocol.

In [7], the XCHARM cross layer routing protocol that chooses the transmission rate by combining the interference, and channel fading. It proposes the inter-channel model that determines the adjacent channel interference; the channel selection and the fading calculation are integrated into the routing protocol. The route is selected based on the channel which gives high data rates and less interference level. The latency of the path is estimated by packet error, contention, forward error correcting codes and the data rate on the selected channels. The route maintenance is proposed to monitor the network performance and trigger the recovery process in case of link failure.

QoS guaranteed intelligent routing using Hybrid PSO-GA [14] integrates the Particle Swarm Optimization (PSO) and Genetic Algorithm. The QoS parameters and interference is converted into penalty functions. The strength of PSO and GA is combined with this approach to get the optimal solution in the search space. The standard velocity and position update rules from the PSO, and crossover, selection processes from the GA are combined for efficient search in the solution space. The fitness function decides the excellence of each particle and it is calculated by summing up the objective and penalty functions. The violation of QoS constraint is modeled as a penalty function and finding the least cost path is considered as an objective function.

LO-PPAOMDV [17] uses cross layer approach to find congestion free route, by collecting information from MAC. The MAC informs unsuccessful communication to the routing layer to identify the congestion. In [9], routing is considered as a multi constraint problem and route is chosen on more than one constraint such as buffer occupancy, energy and hop count. In Wireless Sensor Networks, the nodes are deployed densely; the congestion occurs near the sink node, so a grid based approach [19] identifies the all nodes direction and then applies quorum methods to avoid congestion.

## 3. System Model

### 3.1 Cross Layer approach

There are two types of cross layer approaches: loosely coupled and tightly coupled. The parameters in one layer are cascaded to another layer in the loosely coupled method. For example, the interference level in MAC layer is intimated to the network layer. Two or three layers combined into a single layer in the tightly coupled method. For example channel assignment and routing is optimized into single layer [11].

Most of the current protocols are insufficient for handling the cross layer interaction. Wireless mesh networks need more interaction between the layers, such as MAC and routing layers, routing and transport layers. In this paper the loosely coupled cross layer approach has been used. In our approach, the MAC layer passes the channel interference and congestion information to the network layer, so that the network layer reroute the packet into the congestion free area.

The cross layer based hop-by-hop approach dynamically monitors the status of the link at the MAC layer and the status is updated to the network layer to find the congestion free path. The Figure 1 shows that the interaction and parameter passing between MAC and routing layers. The MAC layer measures the congestion status, on the basis of contending channel interfered with ongoing transmission of neighboring nodes. Our hop-by-hop cross layer approach uses the RTS/CTS protocol for explicit information exchange.

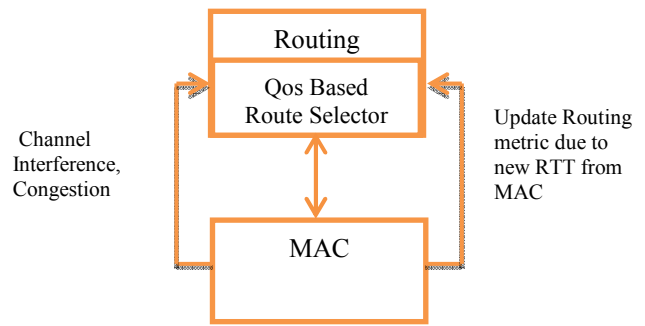

**Figure 1.** Cross layer design.

### 3.2 Model and Motivation

When we drop a ball on the floor, it bounces back, but the ball will not reach its starting position. It's a classic problem in physics. The ball's behavior is identical of a sphere-shaped spring. When the ball hits the floor, it applies a force on the floor greater than its weight, and the floor applies an equal force back. The ball is compressed by this force and the gravitational force. Hooke's law is satisfied for small compression. The gravitational potential energy of the ball before the drop is converted into kinetic energy and eventually into elastic potential energy when the ball is compressed. Some of the energy is converted into thermal energy by internal friction, as the ball is not perfectly elastic. The thermal energy is not converted back. The ball does not reach its initial height, due to its initial gravitational potential energy is converted into thermal energy. We note the phenomenon of "energy loss", characterized by the COR, the ratio of the speed of the ball after bounce to the speed of the ball before bounce. A perfectly hard floor is a stationery floor, incapable of moving itself. The "stationery behavior" is noted, further. The definitions below are significant in the context.

$$Coefficient\ of\ Re\,stitution = \frac{Re\,bound\ \ Speed}{Incidence\ \ Speed} \quad (6)$$

$$\frac{KE_{rebound}}{KE_{rebound}} = \frac{V_{rebound}^2}{V_{rebound}^2} = Coefficient\ of\ Re\,stitution^2 \quad (7)$$

The network is assumed like a gravitational field, the packet is viewed like a ball, moving from source to the gateway, sending a packet and receiving acknowledgement can be viewed as a bouncing ball. The movement of the packet is decided by the gravitational force field.

WMN, nodes are stationary, analogous to the perfectly hard floor. The loss of height could be translated to different path lengths a message may traverse, which is due to the loss of energy explained above. Dynamic routing can be interpreted as energy transfer between nodes, i.e. a persistent interaction among nodes such that messages are transmitted. A good enough measure of energy transfer is explained by kinetic energy, the definition of which is well known.

Let us consider two objects: Object 1 and Object 2, and they are colliding with each other, in this case, the COR is denoted by

$$COR = \frac{(V2 - V1)}{(U1 - U2)} \quad (8)$$

Where:

$V_1$ is the final speed of Object 1 after impact
$V_2$ is the final speed of Object 2 after impact
$U_1$ is the initial speed of Object 1 before impact
$U_2$ is the initial speed of Object 2 before impact.

The COR is considered in evaluating the performance of the proposed approach.

In the proposed approach, each node in the network is assigned with gravitational potential $V(v)$, and the interaction (transmission) between the nodes in its vicinity is influenced by force. Let us assume that the packet $p$ in node $v$ is forwarded to the neighbor node to reach the gateway $g$. The next hop neighbor is identified through the potential field difference between node $v$ and other neighbors.

Assume that $w$ is the neighbor of $v$, here the force is defined as

$$F(v, w) = V(v) - V(w) \quad (9)$$

In this paper, the force is interpreted as delay and the packet $p$ on node $v$ is forwarded to the next hop node which is having a minimum delay or force $F(v, w)$.

If the node $v$ chooses the node $w$ as next hop rather than node u , then it must hold

$$F(v, w) < F(v, u) \quad (10)$$

The coefficient of restitution measures the elasticity of collisions. A perfectly elastic collision has a COR value of 1 and kinetic energy is well-maintained and multi hop transmissions may take place. A perfectly inelastic collision has a COR value of 0. The pair of object with zero COR, stops bouncing at all and it implies no transmission of messages.

### 3.2.1    Length field

The length (distance) is estimated to find the shortest distance between the sender and the gateway. Each packet is transmitted towards the gateway on the basis of the length field. We define the length field as:

$$V_l^{\,g}(v) = length(v) \quad (11)$$

Where $length(v)$ is the length of the node $v$ to the gateway. The $length(v)$ is the shortest path which is calculated by considering the RTT as routing metric, So, $length(v)$ will have a less RTT value. The distance between the node $v$ and the node $u$, specifically $V_l^{\,g}(v, u)$, is represented in ms. The length field $V_l^{\,g}(v)$ is time-based and it dynamically changes when there is any change in the Internet traffic. When the node $v$ has more than one neighbor with different RTT values, then the node $v$ chooses the node with less RTT value as the next hop node. In this fashion, every node calculates the $length(v)$ to discover the list of neighbors towards the gateway, and the nodes maintain a routing table, which contains next hop neighbor and its RTT value. In WMN, redundant paths do exist, so our aim is to consider all the nodes and all the possible routes to discover the congestion free path to route the packets.

### 3.2.2    Modified RTS/CTS mechanism

The objective of finding the congestion status at MAC layer is to avoid the packets moving to the interference area. In this approach, a node selects any of its neighbors to forward the packet by inspecting the channel interference. Specifically, a node selects one of its neighbors with less interference towards the gateway, as a next hop node and it transmits the packets in interference free. Moreover, the congestion at MAC layer is primarily caused by co-channel interference, self-interference and partial channel interference. The congestion status of the link is evaluated based on the RTS/CTS protocol in IEEE 802.11.

In this section, we propose modified RTS/CTS mechanism, and the following assumptions are made:

- The MRMC WMN with 11 channels available for use and the data transmission rate is same for all the channels. Since the channels overlap with each other, transmission in one channel interferes with another channel.
- Each router is equipped with multiple transceivers and assigned to different channels. So the router can simultaneously send and receive on different channels at the same time.

For example, let us denote two nodes: node1 and node2. When node1 has a data to send to node 2, the node1 and node2 exchanges the RTS (Request to Send) and CTS (clear to Send) packets to reserve the idle channel. The preferable channel list (PCL) table is maintained by each node [5] and it contains the list of desirable channels, which helps in avoiding the interference.

The level of preference is divided into three categories:

- **High preference:** The channels that have already been selected by the node in the current beacon interval and each node will have at most one channel is in this state.
- **Medium preference:** The channels that are yet been taken by the node or neighbors within the transmission range of this node.
- **Low preference:** The channels that have already been taken by at least one of its neighbor within the transmission range of this node.

The node1 prepares to send a packet to the node2 and it selects the channel c1. The node1 is configured with channel c1 and it sends RTS packets to node2. The node2 examines the channel c1, to check if any interference with ongoing transmission in node2.

Algorithm 1 describes the RTS/CTS method for QoS guaranteed application. When node1 wants to send a packet to node2, firstly, the node1 has to carefully select a channel which is not interfering with other neighbor nodes. The node1 uses the CSMA/CA to detect the co-channel interference, to identify if the medium is busy, and then the node1 tries with the back-off algorithm. But, the adjacent channel interference is not detected easily and dealing with the same is important as it would decrease the throughput dramatically.

   $i$ - Number of interfaces at node 2.
   $C[i]$ - Assigned channel number at node 2

---

### *Algorithm 1: RTS/CTS for QoS guaranteed Application*

---

*for j=0: i*
  *If c1 equals c[i] then*
      *Defer transmission*
  *else if c[i] equals to channel 1 to 6 then*
    *If (c1 =(c[i] +5)) then*
        *There is no interference and no congestion*
        *in the channel*
        *Send CTS*
    *else*
        *Defer transmission*
  *else if c[i] equals channel 7 to 11then*
    *if ( c1 = =(c[i]+ 5)) mod 11 then*
        *There is no interference and no congestion in*
        *the channel*
        *Send CTS*
    *else*
        *Defer transmission*
  *else*
    *Defer transmission*
  *i ← i+1*
*End for*

---

The node2 has to verify whether c1 is interfering with the channels assigned to other radio. If c1 value is matched with any of its interface channel number, then it is self-interference, so node2 rejects the transmission. If c1 is mutually orthogonal to already assigned channel number in node2, then there is no interference and no congestion in the channel. Hence, the node2 sends CTS to node1.

Algorithm 2 describes RTS/CTS method for delay tolerant application. The channel separation between c1 and other interfaces of node2, and its channel number happens to be 4, and then it is partially overlapping channels in the link with less interference. This is suitable for the application which is capable of tolerating delay and packet drop.

In MAC layer, the logical status of the link is the congestion, but in TCP layer, if the buffer is occupied, then it is regarded as physical congestion. The RTS/CTS exchange eliminates the packet collision due to the channel interference as well as the over saturation of the MAC layer. The performance degradation of TCP in wireless mesh networks is primarily due to contention delay caused by RTS/CTS mechanism.

---

### *Algorithm 2: RTS/CTS for delay tolerant Application*

---

*for j=0: i*
  *If c1 equals c[i] then*
      *Defer transmission*
  *else if c[i] equals to channel 1 to 6 then*
    *If (c1 =(c[i] +4)) then*
        *c1 is overlapping partially, so less*
        *Interference and it is suitable for application*
        *Tolerating packet drop*
        *Send CTS*
    *else*
        *Defer transmission*
  *else if c[i] equals channel 7 to 11then*
    *if ( c1 = =(c[i]+ 4)) mod 11 then*
        *C1 is overlapping partially, so less*
        *Interference and it is suitable for application*
        *tolerating packet drop*
        *Send CTS*
    *else*
        *Defer transmission*
  *else*
    *Defer transmission*
  *i ← i+1*
*End for*

---

### 3.2.3 Cumulative RTT

The delay comprises of three components: propagation delay, transmit delay and queue delay. But in many situations, we are interested in calculating only the total time it takes to transmit a packet from the sender to the receiver and to receive the ACK back. This is regarded as RTT. Assume that a node $v$ receives a data from node $u$, and the node $v$ does not always select the same node to forward the packet. According to the traffic condition, the delay between the two nodes may change dynamically, that result in the same node is not being selected as a next hop.

In this approach, hop-by-hop RTT is estimated, in other words, the delay between the neighboring nodes are individually measured and then cumulative RTT is taken at the sender node. The hop-by-hop delay consists of three components: queue delay, contention delay and transmission delay.

- **Queue delay:** The time interval between the packets reaches the queue and moves to the head of the queue.
- **Contention delay:** The time interval between the packet at the head of the queue and to gain access to the physical channel through the channel access mechanism RTS/CTS.

The queue delay and contention delay are depicted in Figure 2. The contention delay in WMNs with multiple radios is significantly higher compared to the wired network.

*Hop-by-Hop delay= queue delay + contention delay*
              *+transmission delay*          (12)

We have assumed that the packet size is fixed for all the transmission, so the transmission delay does not change dynamically. The queue delay is primarily determined by the contention delay which is the dominant portion of the total hop-by-hop delay.

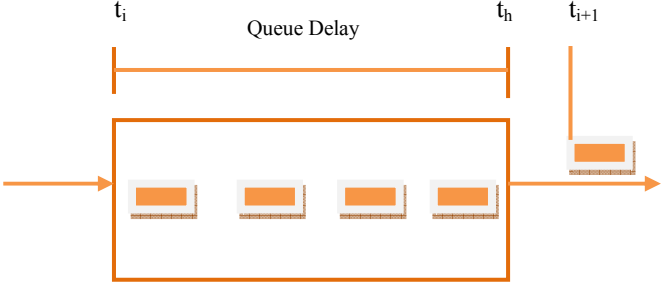

**Figure 2.** Delay in Queue.

For each frame, the variables $t_i$, $t_h$, $t_{i+1}$ are maintained to store time components. The variable $t_i$ is used to hold the arrival time of the frame at node $i$, and $t_h$ records the time at which the frame reaches the head of the queue. The $t_{i+1}$ record the time at which the frame is transmitted to the physical medium of node $i$. The time difference between $t_h$ and $t_i$ is called as queue delay and the time difference between $t_{i+1}$ and $t_h$ gives the contention delay.

$$Queue\ \ delay = t_h - t_i\ \ (13)$$

$$Contention\ \ delay = t_{i+1} - t_h\ \ (14)$$

The function $Q(v)$ describes the queue delay at node $v$. The $Q(v)$ defined as

$$Q(v) = (t_h - t_i) + (t_{i+1} - t_h)\ \ (15)$$

The two potential fields, queue delay and contention delay, are the key features of our approach and are used in making routing decision. For simplicity, queue delay and contention delay are combined linearly as follows:

$$Q(v) = (1 - \alpha)(t_h - t_i) + \alpha(t_{i+1} - t_h)\ \ (16)$$

Where $0 \leq \alpha \leq 1$, if the value of $\alpha$ is zero, then there is no contention delay, and only queue delay at the node. If the value of $\alpha$ is one, then there is no queue delay, but contention delay at the node. If the value lies between zero and one, then both queue and contention delays at the node. The parameter $\alpha$ controls the degree of influence of two potential fields for making routing decision.

Cumulative RTT at node $V$

$$V_P^g(v) = min \sum_0^n RTT(v)\ \ (17)$$

where $\sum_0^n RTT(v)$ is the cumulative RTT from node $v$ towards the gateway $g$. Here $n$ is the number of nodes or hops from source to gateway $g$. The cumulative RTT gives the congestion towards the gateway.

Each node in the network sends a packet to the immediate neighboring node to find out the hop-by-hop RTT and updates its own routing table. The sender node compares RTT value received from all of its neighbors, and chooses the next hop with less RTT value, and then finds the cumulative RTT towards the gateway using the equation 2. Each node in the network recursively doing this process, so, it can determine the congestion and then make a decision to select the next hop.

## 4. Proposed Routing Algorithm

When the node is ready with packets to be sent, it first sends RTS to check if the neighboring node is not congested; in case the neighbor is congested, then the sender waits for some amount of time. Once the sender receives CTS, it starts sending the packets to the neighboring node and subsequently waits for the acknowledgement to calculate the RTT value. The RTT value depends on various factors such as: the rate at which data is transferred from the source, the medium used for the transmission (i.e. a wireless, optical fiber or copper), the distance between the source and neighboring nodes, the presence of noise in the circuit, the number of other requests pending at the intermediate nodes, and the speed at which the intermediate node functions. RTT estimation can be used in routing algorithms for calculating the optimal routes.

For every hop, sampleRTT is calculated by the difference between the packet sent time and ACK received time. The sampleRTT may vary from packet to packet due to dynamic nature of the channel. In order to find out the actual RTT, the average value of sampleRTT is calculated and the AverageRTT [4] is estimated as

$$Difference = sampleRTT - AverageRTT\ \ (18)$$

$$AverageRTT = AverageRTT + (\delta \times Difference)\ \ (19)$$

Where $\delta$ is between 0 and 1.

Since the wireless topology changes dynamically, each node should be able to learn the routes quickly. If any of the nodes are inactive, then the protocol excludes them from the path. So, the hello messages are used by the nodes to indicate activeness and inactiveness to its neighbors. The nodes which are active respond quickly to the new route requests. Hence, there is a need for on-demand routing, which can be achieved using the AODV algorithm.

### 4.1 Routing in typical AODV Approach

The AODV considers the hop count as a routing metric to find the shortest path between the sender and the gateway, which does not account the interference on that path. In order to reduce the interference, AODV chooses the routes by keeping RTT as a metric.

In Figure 3, the mesh topology where the route setup is based on the hop count as a metric, and with a typical AODV approach, the source sends a RREQ to the destination node (Gateway). The route request from node5 reaches the destination node4 through path p1 (5-4) faster than through path p2 (5-7-6-4). Since the number of hops is less in path p1, p1 is selected even though more interference on that link. As the selected channels in path p1 are having a high packet drop, it is necessary to dynamically monitor the delay and accordingly select the path by considering the current channel quality, to reroute the packets.

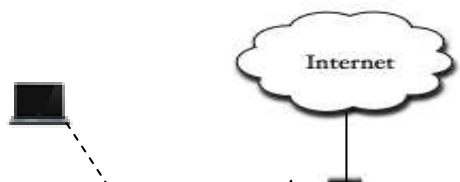

AverageRTT is estimated for each hop in the network. The values of RTT are sorted and the routing table is reconstructed by replacing RTT as its link values. Again, the route discovery module rediscovers the congestion free alternate path and the new throughput is obtained from the network; this new throughput and the older throughput are analyzed to compare the performances.

---

### *Algorithm 3: CoRCiaR Algorithm*

---

*Begin (CoRCiaR Algorithm)*
  *For ( i = 1 to n ) do*
    *If  no route exist then*
       *Perform AODV routing algorithm*
       *Send packets through the shortest path*
    *End if*
    *Calculate the throughput and delay in AODV*
    *For each hop*
      *Estimate RTT*
      *Assign RTT value as a link cost*
    *End for*
    *Select the route with minimum RTT*
    *Calculate the throughput again*
    *Compare the throughput obtained from CoRciaR with AODV routing.*
    *Find the COR values to check the elasticity of the collision and evaluate the performance of the algorithm*
  *End for*
*End for (CoRCiaR Algorithm)*

---

## 5.  Simulation

The performance of CoRCiaR protocol is evaluated using the NS2.29 simulator with MRMC patches included. The simulation uses AODV for dynamic routing and modified RTS/CTS protocol at MAC layer. The nodes are deployed randomly in a 1500 x 800m area for evaluating the performance in chain and random topologies. In random topology as the name suggests, the distance between the nodes are random, wherein the chain topology has the fixed distance of 150m between the nodes. The transmission range is set to 250m, and the interference range is set to 550m. The default data rate 1Mbps is used and the packet size is set to 1000bytes. The traffic types used in the simulation is TCP and the simulation was performed for 500s. The comparison study is performed between the CoRCiaR with TCP-AP (TCP with Adaptive Pacing) [2], and semi-TCP with ACK [11]. For simulation, the network is organized with 20,40,60,80 and 100 nodes, randomly distributed in a flat grid area.

**Figure 3.**  Routing in Mesh Architecture.

### 4.2 Routing in CoRCiaR Approach

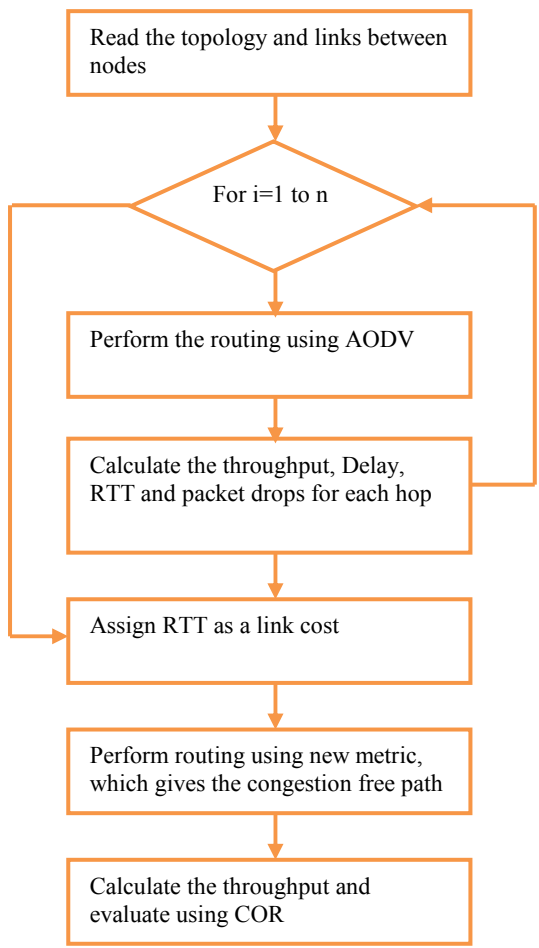

**Figure 4.** Routing using RTT as a metric

Figure 4 explains how the CoRCiaR protocol is performed using RTT as a metric. Initially the route discovery module finds the shortest path based on the number of hops between the source and the gateway. Packets are sent through the shortest path using the typical AODV algorithm and then

**Table 1.** Simulation Settings.

| Parameters | Values |
|---|---|
| Platform | NS2 version 2.29 with MRMC patch |

| Network Area | 1500m X 800m |
|---|---|
| Propagation model | Two ray ground model |
| Network Topologies | Chain topology and Random topology |
| Transmission Range | 250m |
| Interference Range | 550m |
| Frequency | 2.4GHz |
| Traffic Type | TCP |
| Channels | 1-11 |
| Packet Size | 1000bytes |
| Maximum queue length | 50 |
| Simulation Time | 100s |
| Transport Type | TCP |
| Data Rate | 1 Mpbs |

### 5.1 Evaluation Criteria

- **Throughput:** The throughput is measured at the gateway, and it is obtained by averaging out all the flows at a given time.

$$Throughput = \frac{Number\ of\ packets\ successful\ ly\ received\ by\ the\ gateway}{Number\ of\ packets\ successful\ ly\ sent\ by\ the\ source}$$

- **End-to-End Delay:** The cumulative measure of delay, the packet to traverse, from source to destination nodes. It includes queue, propagation and transmission delays.
- **RTT:** It is the time taken by a packet to reach destination plus ACK back to the source node.
- **COR:** It is the ratio of throughput, before and after the collision at MAC layer.

$$COR = \frac{Throughput\ after\ drop\ due\ to\ collision\ at\ MAC}{Throughput\ before\ impact\ of\ collision\ at\ MAC}$$

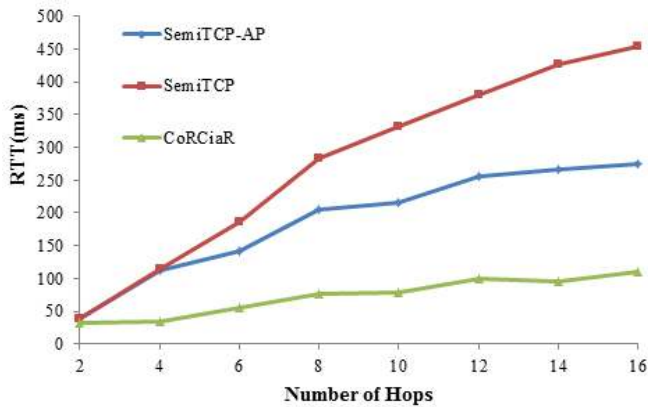

**Figure 5.** RTT against Number of Hops.

To analyze the performance of the routing algorithm, the simulation of two existing congestion control methods, semi-TCP with ACK, and TCP-AP, were performed. From Figure 5, it is evident that the increase in the path length i.e. number of hops, also increases the RTT values. The Figure 5 depicts RTT values for all the three schemes; the x-axis denotes the number of hops, while the y-axis denotes RTT values in milliseconds. The graph shows that the proposed scheme outperforms other two approaches with the clear advantage of predicting the traffic condition and interference at each hop. In the conditions like nodes deployed at random fashion and the network with high interference, the proposed method yields significantly less delay.

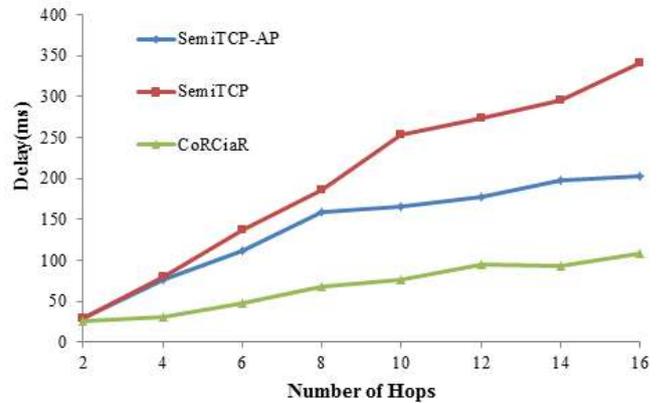

**Figure 6.** Delay Vs Number of Hops.

From figure6, it can be observed that the proposed method drastically reduces the packet delay compared to the other two methods.

Figure7 show cases the throughput obtained by semi-TCP with ACK, TCP-AP and CoRCiaR. The CoRCiaR performs well even with an increased number of hops. The throughput decreases dramatically when the number of hop increases and this is due to channel sharing in the MAC layer. The throughput of CoRCiaR protocol is stable, when the number of hops reaches 4 or more. The other two algorithms obtained lower throughput than the CoRCiaR protocol as the number of hop increases.

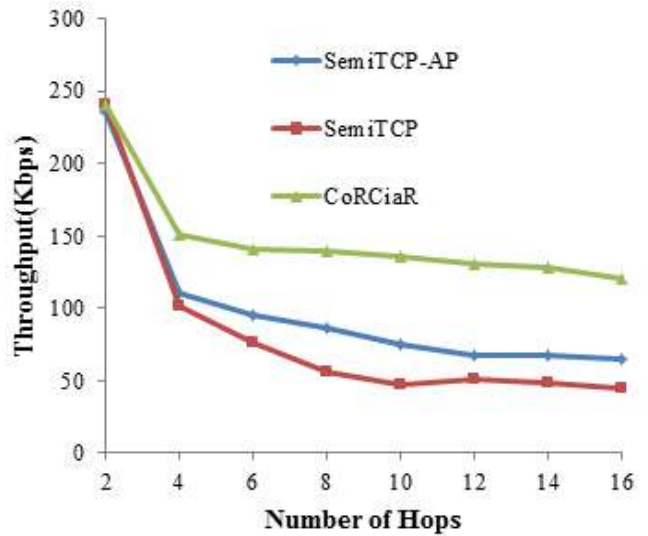

**Figure 7.** Throughput Vs Number of Hops.

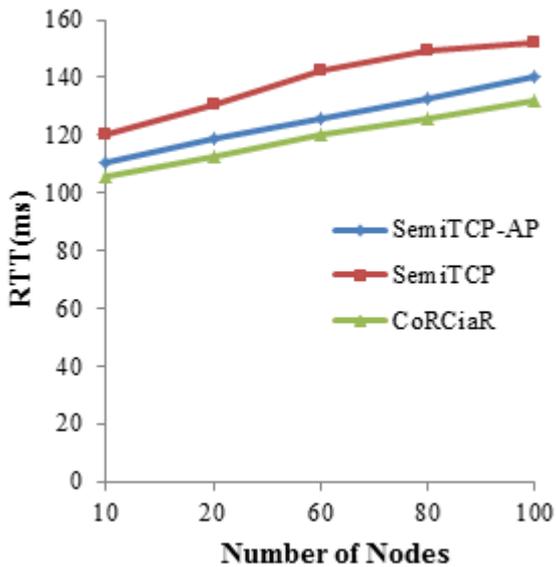

**Figure 8.** RTT Vs Number of Nodes.

RTT is increased when the number of nodes deployed in the network is high. Figure 8 shows that CoRCiaR protocol gives less RTT value compared to SemiTCP and SemiTCP-AP. Figure 9 shows that the throughput of our protocol is higher than the other two methods.

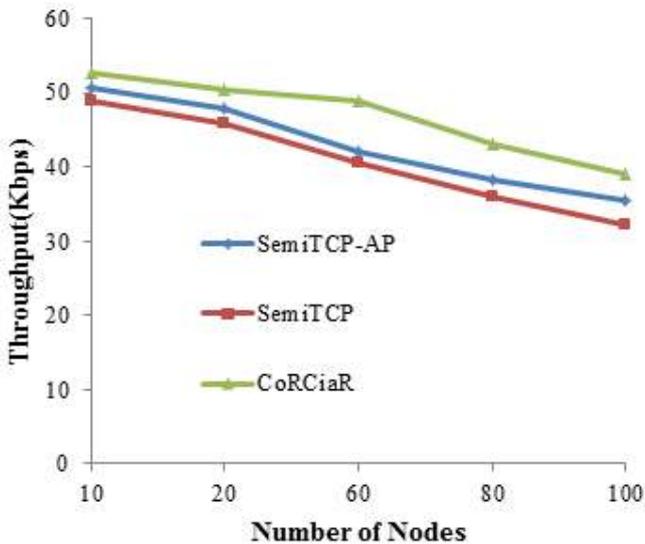

**Figure 9.** Throughput Vs Number of Nodes.

### 5.2 Performance evaluation using COR

In wireless networks, throughput depends upon the packet drop and the whole network performance is determined by the throughput which is calculated using the COR. The COR is the ratio between the throughput derived using typical AODV and the throughput derived through our approach. The throughput is inversely proportional to the RTT value.

The COR values lie between zero and one, indicates the elasticity of the collision. If the COR value is 1, then no packet drops in the network and this condition are known as perfectly elastic collision, which produces the consistent improvement of throughput in the network. If the COR value is 0, then significant packets have been dropped and this is known as inelastic collision, in which the performance consistently decreases. When the COR value ranges between

0.0 and 1.0 , few packets drop are seen in the network, which results in consistent improvement of throughput in the network and the same is called as partially elastic collision. Table 2 shows the throughput of SemiTCP, CoRCiaR and COR values. It indicates that the proposed method produces the higher throughput than the other two algorithms. So, the COR values are used to evaluate the performance of the network depending upon the values.

**Table 2.** COR values.

| Throughput of SemiTCP with ACK(Kbps) | Throughput of CoRCiaR (Kbps) | COR |
|---|---|---|
| 483.133 | 483.133 | 1 |
| 240.936 | 240.936 | 1 |
| 154.658 | 177.829 | 0.869701 |
| 101.137 | 150.523 | 0.671904 |
| 84.6593 | 147.935 | 0.572274 |
| 75.6836 | 140.5726 | 0.538395 |
| 57.1282 | 140.5449 | 0.406477 |
| 56.5467 | 139.8836 | 0.404241 |
| 47.2099 | 138.7724 | 0.340197 |
| 47.5675 | 135.6574 | 0.350644 |
| 48.9261 | 133.5736 | 0.366286 |
| 0.8329 | 130.2198 | 0.390362 |
| 48.2715 | 129.3132 | 0.373291 |
| 48.1595 | 128.3545 | 0.375207 |
| 47.8169 | 122.8472 | 0.389239 |
| 45.1668 | 120.2656 | 0.375559 |
| 48.5566 | 118.7433 | 0.408921 |
| 46.7605 | 117.8355 | 0.396829 |
| 49.2422 | 115.1323 | 0.427701 |

## 6. Conclusion

WMN is considered as one of the most reliable and low cost network to provide broadband Internet access. The congestion control in MRMC WMN is different from the traditional congestion control. In this paper, we have proposed the CoRCiaR protocol to reroute the traffic in the congestion free path in WMN and the RTT in each hop is considered in making the routing decision. In multi-channel, the adjacent channel interference is very severe, so there would be a significant amount of packet loss and that results in performance degradation. Some modifications in the RTS/CTS scheme can significantly improve the throughput. The proposed method decreases the packet drop, packet retransmission and end-to end delay. Simulation results clearly demonstrate that the proposed scheme increases the network performance compared to other methods like semi-TCP, TCP-AP.

COR describes the inelasticity of the collision which measures the performance of the network and also useful to make routing decision on the multi hop environment.

The benefits of CoRCiaR protocol are:
- The traffic is distributed across all the 11 channels.
- The reliability and connectivity are sustained in WMN.

- Broadcasting and multicasting capabilities due to multiple channels.